\documentclass{webofc}
\usepackage[varg]{txfonts}   % Web of Conferences font
\def\snn{\mbox{$\sqrt{s_{_{NN}}}$}}
\begin{document}
\title{Simulating collectivity in dense baryon matter with multiple fluids}
%\subtitle{MUFFIN: Three-fluid hydrodynamics}

\author{\firstname{Jakub} \lastname{Cimerman}\inst{1,2} \and 
	\firstname{Iurii} \lastname{Karpenko}\inst{1}%\fnsep\thanks{\email{iurii.karpenko@cvut.cz}} 
	\and
	\firstname{Boris} \lastname{Tom\'a\v{s}ik}\inst{1,2}\fnsep\thanks{\email{boris.tomasik@cern.ch}} 
	\and
        \firstname{Pasi} \lastname{Huovinen}\inst{3}%\fnsep\thanks{\email{pasi.huovinen@uwr.edu.pl}}
        % etc.
}

\institute{Fakulta jadern\'a a fyzik\'aln\v{e} in\v{z}en\'yrsk\'a, \v{C}esk\'e vysok\'e u\v{c}en\'i technick\'e v Praze, Praha, Czechia
\and
Univerzita Mateja Bela, Bansk\'a Bystrica, Slovakia
\and
Incubator of Scientific Excellence---Centre for Simulations of Superdense Fluids, University of Wroc\l{}aw,  Poland
          }

\abstract{%
A novel three-fluid dynamical model for simulations of heavy-ion collisions at RHIC Beam Energy Scan programme, called MUFFIN, has been developed. The novelty consists of modular inclusion of the Equation of State, use of hyperbolic coordinates that allow to simulate higher collision energies, and fluctuating initial conditions. The model reproduces rapidity and $p_t$ spectra of collisions from $\snn=7.7$ to 62.4~GeV. Ideal fluid is assumed, and the elliptic flow is over-predicted, particularly at lower collision energies. 
}
\maketitle
\section{Introduction}
\label{s:intro}

A suitable simulation tool is necessary for theoretical understanding of nuclear collisions at energies of the RHIC Beam Energy Scan (BES). Strongly interacting matter produced there is anticipated to pass states in vicinity of a critical point on its phase diagram. The bulk matter is directly treated in fluid-dynamical approaches. Such simulations, embedded into hybrid models, are very successful in describing evolution and many observables from nuclear collisions at highest available energies at the LHC and RHIC. In a hybrid model, there is usually a module that models the initial deposition of energy and thus sets the initial conditions for a hydrodynamic code which follows. This is appropriate because the time of establishing the initial conditions  is an order of magnitude shorter than the subsequent evolution. 

Simulations at lower energies are more complicated, though. Firstly, the interpenetration time of the two nuclei is of the same order as the whole subsequent lifetime of the fireball.  
Secondly, the bulk is not baryon free, unlike at high energies,  and the flow of baryon number must be treated explicitly. While the latter issue is addressed  by an addition of the corresponding conservation equation, the former requires a revision of the initial conditions. One could use some model to parametrise the initial conditions, but they would have to be applied rather late, when the nuclear passage has been completed, already. Hence, the fluid-dynamical simulation would miss a large part of the most interesting dense period of the collision. 

Two solutions could be envisaged. Firstly, one can proceed via so-called dynamical initialisation:  a transport code is run and the space is coarse-grained. In each time step, cells that contain energy above some critical value are turned into fluid which carries the same energy and momentum. The question is, however, how to treat both fluidised and non-fluidised cells in the simulation concurrently.  The second possibility is to describe the incoming nuclei as two fluids and the production of the hot matter as a result of friction between these fluids. This is the multi-fluid dynamic model. Here, we introduce a novel three-fluid dynamic model, which we call MUFFIN, for MUlti Fluid simulation for Fast IoN collisions \cite{Cimerman:2023hjw}.

%%%%%%%%%%%%%%%%%%%%%%%%%%%%%%%%%%%%%%%%%%%%%%%%%%%%%%%%%%%%%%%%%%%%
\section{A new three-fluid model}

Three-fluid dynamic models have been constructed in the past \cite{Csernai:1982zz,Mishustin:1991sp,Ivanov:2005yw}. We base MUFFIN on \cite{Ivanov:2005yw}, but develop the model from scratch and implement novelties: i) sampling of the initial state that allows to simulate event-by-event fluctuations; ii) Equation of State (EoS) implemented as a module that can be easily exchanged; iii) use of hyperbolic coordinates $\tau=\sqrt{t^2-z^2}$ and $\eta = \frac{1}{2}\ln ((t+z)/(t-z))$ which allows  to use the model at higher collision energies and  to apply it for the whole range of the RHIC BES energies. 

\begin{figure}[t]
\centering
\sidecaption
\includegraphics[width=0.45\textwidth]{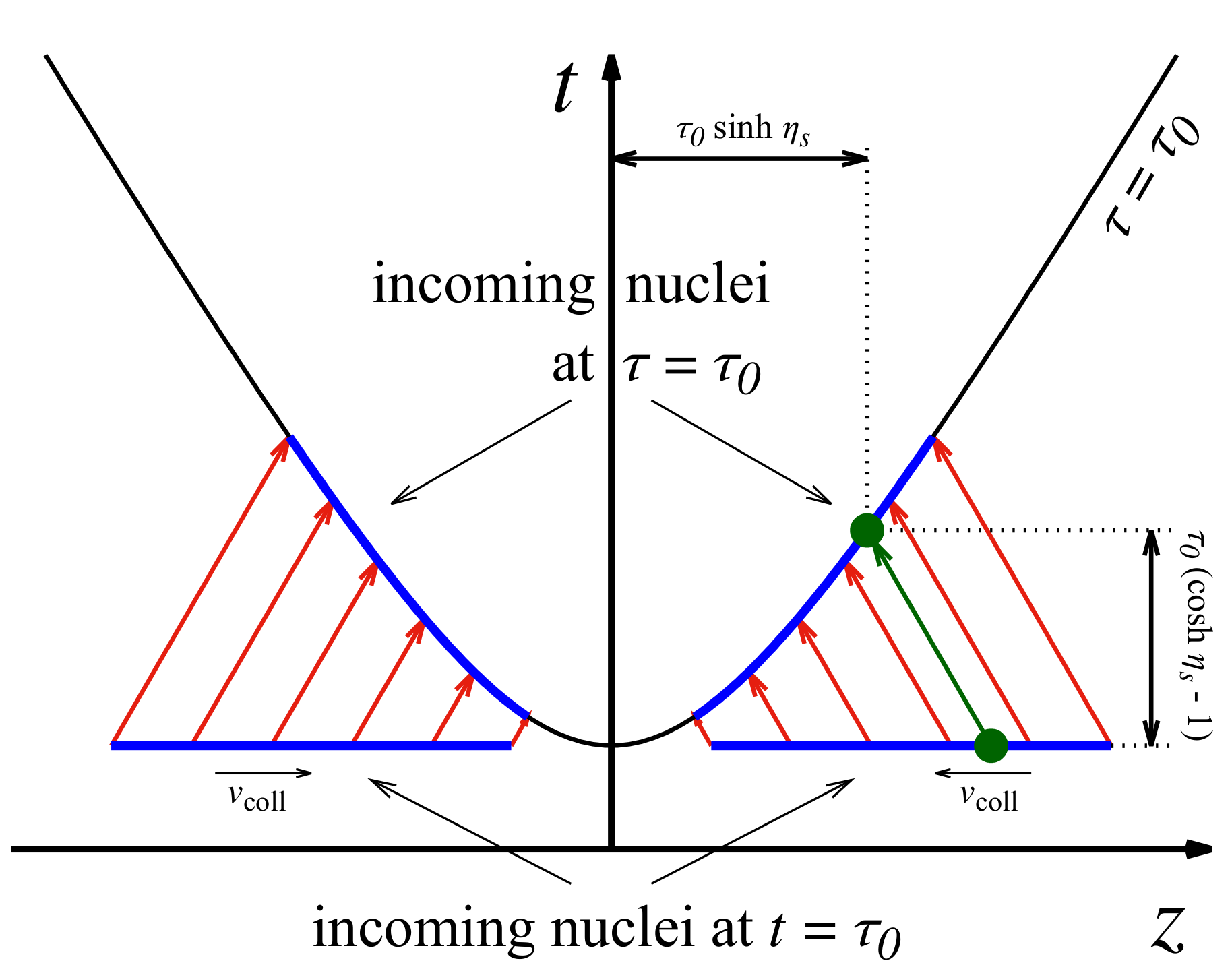}
\caption{Construction of the initial conditions. Nucleons within the incoming nuclei are first sampled at the fixed global time $\tau_0$ so that the two nuclei still do not overlap. Then, they are projected and smeared along the $\tau=\tau_0$ hyperbola, along which the initial conditions for hydrodynamics in hyperbolic coordinates are formulated. }
\label{f:ic_diag}       % Give a unique label
\end{figure}

Construction of the initial conditions is illustrated in Fig.~\ref{f:ic_diag}.
%%%%%%%%%%%%%%%%%%
%%%%%%%%%%%%%%%%%%
Nucleons are first placed at the hypersurface with equal centre-of-mass (CMS) time $t = \tau_0$. They are sampled according to Woods-Saxon distribution with  $z$-coordinates  appropriately Lorentz  contracted. The positions of the whole nuclei are chosen such that they do not yet overlap, i.e., they are still before the collision. The code, however, is formulated in hyperbolic coordinates. Hence, the nucleons are then projected on the $\tau=\tau_0$ hyperbola. Since we are still before the collision, they are propagated along worldlines that correspond to free motion of one or the other nucleus. Moreover, as the earliest global time on the hyperbola is $\tau_0$, and any light-cone starting on the hyperbola lays completely above it, we are guaranteed that the nuclei projected in such a way will not have interacted, yet. Note, that $\tau_0$ is just a technical parameter, which actually determines the curvature of the initial proper time hyperbola, since the procedure always starts with two nuclei placed at CMS time $t=\tau_0$.  To serve as initial conditions, the positions of the nucleons along the hyperbola are smeared when determining the resulting energy, momentum, and baryon densities \cite{Oliinychenko:2015lva}.

For the fluid dynamical evolution, vHLLE package is used \cite{Karpenko:2013wva}.  When in contact, the two fluids that correspond to the incoming nuclei, produce a third fluid which simulates the created hot matter. Each of the fluids is evolved separately, even though they exist at the same place. The motivation for such a treatment is in the distribution of  matter, participating in the collision, in momentum space, or even better, in rapidity. The two incoming nuclei occupy two  narrow peaks at forward and backward rapidities. They will be slowed down  and widened during the collision. The newly produced matter will occupy some interval around midrapidity. Moreover, energy and the baryon number may be deposited from the incoming nuclei to the fireball matter differently. The kinematic distinction of the three different components leads to the idea of the three-fluid dynamics. At lower energies the matter of the incoming nuclei may be slowed down to such an extent that its velocity is very similar to that of the fireball, but this does not invalidate our motivation to treat them as separate fluids. 

Technically, energy and momentum are transferred from one fluid to another in a process of friction. Hence, conservation law does hot hold for each fluid separately, but for all three of them in total
\begin{equation}
\partial_\mu \left ( T_p^{\mu\nu} + T_t^{\mu\nu} + T_f^{\mu\nu}\right ) = 0\,  ,
\label{e:econ}
\end{equation}
where $T^{\mu\nu}$ are energy-momentum tensors and subscripts $p$, $t$, $f$ refer to projectile, target, and fireball, respectively. 
In the current version, baryon number stays with the projectile and target fluid. For the detailed formulation of the friction terms we refer to \cite{Ivanov:2005yw,Cimerman:2023hjw}.

As the energy density drops due to expansion, fluid-dynamic description is no longer appropriate and particlisation procedure is invoked. The energy flow defines fluid velocity, in the Landau convention.  However, in a three-fluid picture this is ambiguous, since we could define three different flow velocities and the corresponding energy densities, even though none of them would  include all relevant energy.  Therefore, we add together all three energy-momentum tensors from eq.~(\ref{e:econ}) and determine the eigenvalues of the summed tensor. The highest eigenvalue is the energy density. Where its value meets 0.5~GeV/fm$^3$, the freeze-out hypersurface is constructed with the help of the CORNELIUS routine \cite{Huovinen:2012is}. HadronSampler from the SMASH package is run then and the produced hadrons are further propagated in the microscopic transport model SMASH \cite{SMASH:2016zqf}. By construction, the hypersurface is thus common for all three fluids, however, they are all turned into particles separately.

%%%%%%%%%%%%%%%%%%
\begin{figure}[t]
\centering
\sidecaption
\includegraphics[width=0.45\textwidth]{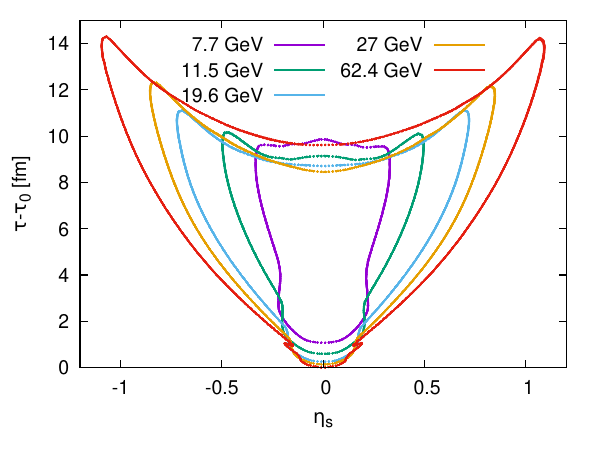}
\caption{The hypersurfaces with energy density at which particlisation happens, as determined by CORNELIUS, for different collision energies. The parts where energy flows into the enclosed four-volume are excluded from particlisation.
}
\label{f:fohs}       % Give a unique label
\end{figure}
%%%%%%%%%%%%%%%%%%
The  hypersurfaces determined in this way are closed, as seen in Fig.~\ref{f:fohs}. At the beginning, the matter of the incoming nuclei enters the collision. In order to disregard such parts of the hypersurface, we require positive outward flow of energy through the freeze-out hypersurface. 

Finally, hadrons are fed into SMASH \cite{SMASH:2016zqf} and evolved within this transport model.

%%%%%%%%%%%%%%%%%%%%%%%%%%%%%%%%%%%%%%%%%%%%%%%%%%%%%%%

\section{Sample results}

With the model, we have calculated the basic observables: rapidity distributions of charged hadrons and net protons, as well as transverse momentum spectra of identified hadrons, which we have reproduced.  For each energy of the RHIC BES programme from $\snn=7.7$ to 62.4~GeV we have simulated 3000 hydrodynamical events, each one of them oversampled with 500 transport events. The model was calibrated  mainly by tuning the friction terms that govern the energy transfer between the fluids. 

\begin{figure}[t]
\centering
\includegraphics[width=0.4\textwidth]{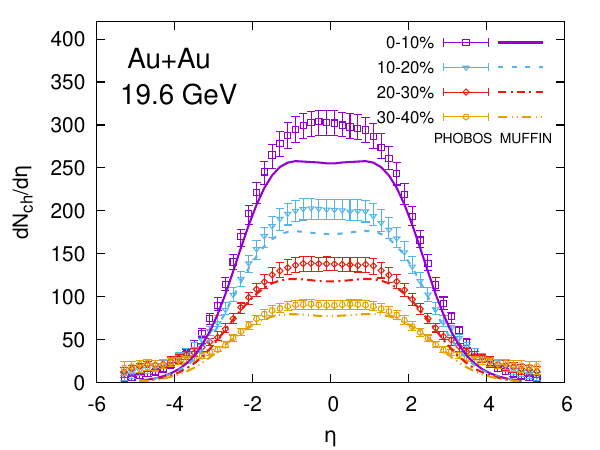}
\includegraphics[width=0.4\textwidth]{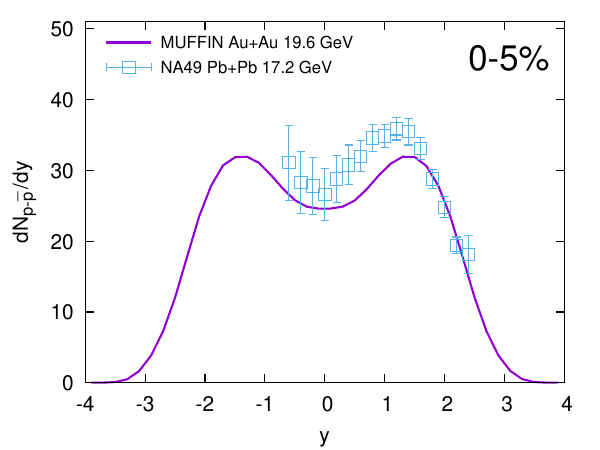}
\caption{Left: pseudorapidity spectra of charged hadrons at $\snn=19.6$~GeV for different centralities, compared to data \cite{PHOBOS:2010eyu}. Right: rapidity spectrum of $p\!-\!\bar p$ in central collisions at 19.6~GeV (calculation) and 17.2~GeV (data NA49 \cite{NA49:1998gaz}).}
\label{f:rs}       % Give a unique label
\end{figure}

For illustration, we show in Fig.~\ref{f:rs} the (pseudo)rapidity distributions for energies $\snn=19.6$ and 17.2~GeV. 
%%%%%%%%%%%%%%%%%%%%
%%%%%%%%%%%%%%%%%%%%
We also reasonably reproduce the transverse momentum spectra, with some tension for antibaryons at lower collision energies \cite{Cimerman:2023hjw}. 

Our model overshoots the elliptic flow depending on centrality and $p_t$, at lower collision energies. Nevertheless, viscosity has not yet been implemented and it is known to reduce $v_2$.

%%%%%%%%%%%%%%%%%%%%%%%%%%%%%%%%%%%%%%%%%%%%%%%%%%%%%%%
 
 \section{Conclusions}

The hybrid model with three-fluid dynamical treatment of the hottest collision phase is a possible way of simulating heavy-ion collisions from the RHIC BES programme. We have constructed a novel implementation of this idea, which includes several modern features and technical advantages. Our use of hyperbolic coordinates allows simulations in a broader range of collision energies. The influence of Equation of State can be studied more easily, as they are inserted as a plug-in module. Event-by-event fluctuations can be investigated thanks to  our construction of the initial state. 

The next step is the introduction of the viscosity and revision of the friction terms so that the (anti)baryons at lower energies and the elliptic flow are reproduced better.

%%%%%%%%%%%%%%%%%%%%%%%%%%%%%%%%%%%%%%%%%%%%%%%%%%%%%%%

\paragraph{Acknowledgements}
This work has been supported by the Czech Science Foundation under No. 22-25026S. BT also acknowledges support by VEGA  1/0521/22.
PH was supported by the program Excellence Initiative--Research University of the University of Wroc\l{}aw of the Ministry of Education and Science.

%%%%%%%%%%%%%%%%%%%%%%%%%%%%%%%%%%%%%%%%%%%%%%%%%%%%%%%

\end{document}